\documentclass[aps,prl,twocolumn,showpacs,amsmath,amssymb,preprintnumbers,10pt,footinbib]{revtex4-1}
\usepackage{graphicx}  
\usepackage{dcolumn}   
\usepackage{bm}        
\usepackage{xcolor}
\usepackage{hyperref}
\usepackage{mathtools}

\usepackage{amssymb,amsmath,graphics}   
\newcommand{\eq}{\begin{equation}}
\newcommand{\eqe}{\end{equation}}
\newcommand{\eqa}{\begin{eqnarray}}
\newcommand{\eqae}{\end{eqnarray}}

\usepackage{epsfig}
\usepackage{amsfonts}
\usepackage{amssymb}
\usepackage{enumitem}
\usepackage{amsthm}
\usepackage{subfig}
\usepackage{extarrows}
\usepackage{bm}
\usepackage{hyperref}
\usepackage{mathrsfs}
\usepackage{bbm}
\usepackage{cancel}

\begin{document}
\title{Hidden zeros are equivalent to enhanced ultraviolet scaling and lead to unique amplitudes in Tr($\phi^3$) theory}
\author{Laurentiu Rodina}
\email{laurentiu.rodina@gmail.com}
\affiliation{Beijing Institute of Mathematical Sciences and Applications (BIMSA), Beijing, 101408, China}

\begin{abstract}
We investigate the hidden amplitude zeros discovered by Arkani-Hamed et al., which describe a non-trivial vanishing of scattering amplitudes on special external kinematics. We first prove that every type of hidden zero is equivalent to what we call a "subset" enhanced  scaling under  Britto-Cachazo-Feng-Witten shifts, for any rational function built from planar Lorentz invariants $X_{ij}{=}(p_i{+}p_{i+1}{+}\ldots{+}p_{j-1})^2$. This directly applies to Tr($\phi^3$), non-linear sigma model, or Yang-Mills-scalar amplitudes, revealing a novel type of enhanced UV scaling in these theories. We also use this observation to prove the conjecture that Tr($\phi^3$) amplitudes are uniquely fixed by the zeros, up to an overall normalization, when assuming an ordered and local propagator structure and trivial numerators. In this context, unitarity (residue factorization) may be viewed as a consequence of the zeros. For Yang-Mills theory, we conjecture the zeros, combined with the Bern-Carrasco-Johansson color-kinematic duality in the form of amplitude relations, uniquely fix the $\lfloor n/2\rfloor$ distinct polarization structures of $n$-point gluon amplitudes. Our approach opens a new avenue for understanding previous similar uniqueness results, and also extending them beyond tree level for the first time.
\end{abstract}

\maketitle
\section{Introduction}
In the modern on-shell bootstrap program, scattering amplitudes of various theories were shown to be constructible via completely new methods, including recursion relations \cite{Cachazo:2004kj,Britto:2005fq,Cheung:2015cba}, Pfaffians \cite{Cachazo:2013hca}, geometrical/combinatorial objects like the Amplituhedron, associahedron, or permutahedron \cite{Arkani-Hamed:2013jha,Arkani-Hamed:2017mur,Cao:2022vou}, Hopf algebras \cite{Brandhuber:2021bsf,Chen:2024gkj}, the double copy procedure \cite{Bern:2010ue}, transmutation operators \cite{Cheung:2017ems}, ansatz based methods \cite{Cheung:2014dqa,Cheung:2016drk}, and most recently from curve counting \cite{Arkani-Hamed:2023lbd}, among many others. These novel perspectives have revealed surprising structures and immense computational simplifications that are completely hidden by the Lagrangian formalism. It is especially shocking that the latest discovery is perhaps the simplest one yet:  termed hidden zeros, these imply that amplitudes vanish for particular kinematic configurations of the external data, and that behavior near these zeros leads to a novel factorization-type property, called splitting \cite{Arkani-Hamed:2023swr} (see also \cite{Arkani-Hamed:2024nhp,Arkani-Hamed:2024yvu,Cao:2024gln,Arkani-Hamed:2024fyd,Cao:2024qpp,De:2024wsy}, \cite{Bartsch:2024amu,Li:2024qfp} for zeros and the double copy, as well as \cite{Cachazo:2021wsz} for early observation of splitting, and \cite{osti_4736008} for very early work on zeros in string theory). Furthermore, in \cite{Arkani-Hamed:2023swr} it was conjectured that these zeros are sufficient to uniquely determine amplitudes in Tr$(\phi^3)$ and non-linear sigma model (NLSM) theories. In this Letter we will prove this conjecture for Tr$(\phi^3)$, establishing an alternative definition of such amplitudes, not based on Feynman diagrams, but on a zero-based bootstrap.

This latter observation about uniqueness ties in with a long list of other uniqueness results \cite{Arkani-Hamed:2016rak,Carrasco:2019qwr,Rodina:2016mbk,Rodina:2016jyz,Rodina:2018pcb,Rodina:2020jlw}, which showed that scattering amplitudes in scalar, gauge, and gravity theories are fully determined by various sets of principles (ranging from gauge invariance to IR or UV scaling), with unitarity, and in some cases locality, being emergent properties. It is highly surprising there is still room for yet new defining properties  in this already very crowded space, suggesting some of these principles may be secretly equivalent to each other.

In this Letter we prove that amplitude zeros are in fact equivalent to a novel "secret" UV scaling under non-adjacent Britto-Cachazo-Feng-Witten (BCFW) shifts \cite{Britto:2005fq}. This equivalence, concretely explained in eq.(\ref{c1}), is valid for all zero types, and for any function built purely from planar invariants, which includes amplitudes (and ansatze) for Tr($\phi^3$), NLSM, and Yang-Mills-scalar (YMs). 

Moving on to pure YM amplitudes, we make another non-trivial observation that, coupled with the Bern-Carrasco-Johansson (BCJ) color-kinematic duality, which enforces linear relations between amplitudes of different external ordering, the zeros uniquely determine all different polarization structures, as classified by the number of dot products of the type $e{\cdot}e$ and $e{\cdot}p$.

\section{Review of amplitude properties}
In this section we briefly review some of the key amplitude properties that we will encounter.

\begin{figure}[htbp] 
   \centering
   \includegraphics[width=2.8in]{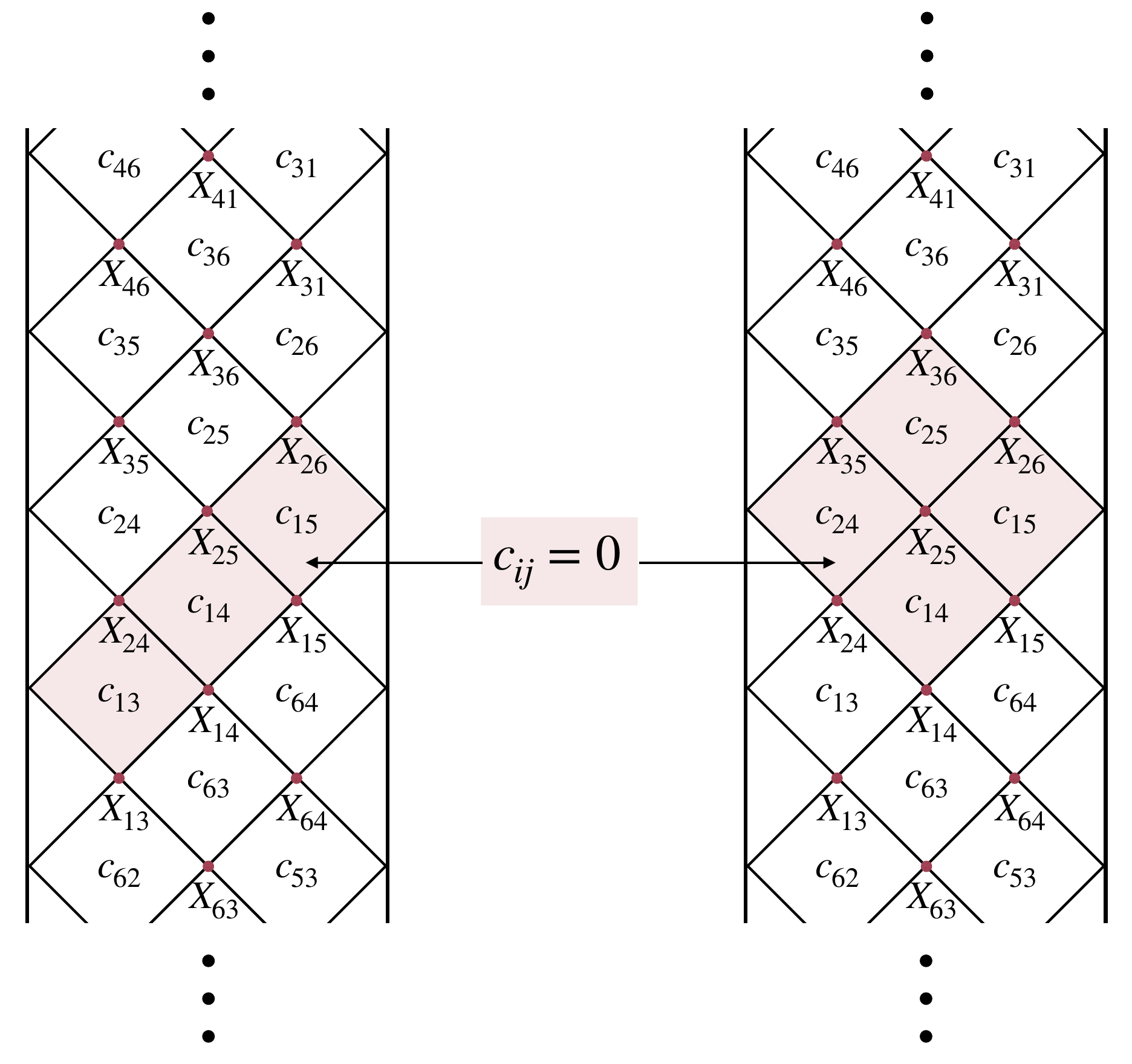} 
   \caption{The kinematic mesh organizes the $X_{ij}$ located at the vertices, and the $c_{ij}$ represented by the squares of a lattice. Above are two 6-point meshes. Note on the boundary of the mesh $X_{i,i+1}=0$, and are omitted. The identity in eq.(\ref{cij}) can easily be visualized as relating the labels of a square and its vertices. Highlighted above are the sets of $c_{ij}$ in a 1-zero and a 2-zero respectively.}   \label{fig:1}
\end{figure}

\textbf{Kinematic data} Tree amplitudes are rational functions of Lorentz invariant dot products of momenta. Typical notations used are the Mandelstam invariants $s_{ij\ldots k}{=}(p_i{+}p_j{+}\ldots{+}p_k)^2$, the planar invariants $X_{ij}{=}s_{i\ldots j-1}{=}(p_i{+}p_{i+1}{+}\ldots{+}p_{j-1})^2$, and the non-planar invariants $c_{ij}{=}{-}s_{ij}$, which will be explicitly reserved for $i,j$ non-adjacent. Here $p_i$ is the $D$-dimensional momenta of particle $i$, subjected to momentum conservation $\sum_i p_i{=}0$ (which implies $X_{ij}{=}X_{ji}$), and the on-shell condition for (in our case) massless particles $p_i^2{=}0$ (which implies $X_{i,i+1}{=}0$). At $n$-points, there is a basis of $n(n{-}3)/2$ kinematic invariants, which we can take as the $X_{ij}$. A crucial relation between the planar $X_{ij}$ and the $c_{ij}$ is given by 
\eq\label{cij}
c_{ij}=X_{i,j}+X_{i+1,j+1}-X_{i+1,j}-X_{i,j+1}\,,
\eqe
to which we will often refer back. The kinematic data is neatly represented graphically in the kinematic mesh \cite{Arkani-Hamed:2023swr}, shown in Figure \ref{fig:1}.

This organization also shows the amplitude zeros, which always set particular non-planar $c_{ij}$ to zero, as we will see later. 

 \textbf{Pole structure and ordered amplitudes} In this paper, we will only consider amplitudes of colored particles, which can be written as a sum over different color-ordered components \eq
\mathcal{A}_n = \sum_{\sigma \in S_n / Z_n} \text{Tr}\left( T^{a_{\sigma(1)}}  \dots T^{a_{\sigma(n)}} \right) A_n(\sigma(1), \dots, \sigma(n))\,.
 \eqe
 In this case, the "partial" amplitudes $A(1,2,\ldots,n)$ have fixed ordering, and we assume all singularities must be of the form $1/X_{ij}$, ie. sums of consecutive momenta squared. We will always use this assumption throughout the paper.\\

 \textbf{Locality} If the singularities of a function correspond to propagators of tree graphs (with either cubic vertices for Tr($\phi^3$) and YM, or quartic vertices for NLSM), we call the function local, otherwise we call it non-local. For instance $1/{s_{12}s_{123}}$ is a local term, while $1/{s_{12}^2}$ or $1/{s_{12}s_{23}}$ are non-local.\\

\textbf{Unitarity} From the optical theorem, unitarity implies factorization of residues of the above poles. Near a pole $P_i^2{=}0$, the amplitude factorizes into two lower point amplitudes
\eq\label{res}
 \mathcal{A}_n({P_i^2=0})\rightarrow \frac{1}{P_i^2}\times A^\textrm{left}\times A^\textrm{right}\,.
\eqe
In our ansatze we will not assume unitarity, but instead we will show it  follows automatically from uniqueness, up to an overall normalization.  \\

\textbf{UV/BCFW scaling}
The renowned BCFW recursion enabled the computation of scattering amplitudes from residues using the Cauchy theorem, applying unitarity as given in eq.(\ref{res}). The poles are accessed via a complex parameter $z$ introduced through a BCFW shift, 
\eq
p_i\rightarrow p_i+z q, \quad p_j\rightarrow p_j-z q\,,
\eqe 
with $p_i{\cdot} q{=}p_j{\cdot} q{=}q^2{=}0$. Crucially, the recursion requires the contour at infinity to vanish, which means that under a BCFW shift, the amplitude must behave as $1/z$ or better at large $z$. Naive power counting shows that individual Feynman diagrams in most theories are badly divergent, but miraculously they combine to produce a so called "enhanced" BCFW (also called UV) behavior, such as $1/z$ for YM and $1/z^2$ for GR,  enabling the recursion in these theories. \\

\textbf{Amplitude zeros}
In \cite{Arkani-Hamed:2023swr}, it was proposed that ordered amplitudes for $\mathrm{Tr}(\phi^3)$, NLSM, and YM vanish when particular kinematic invariants are set to zero. There are different types of zeros, classified by the number of invariants involved. In this paper we will mostly focus on the following so-called "skinny" zero, which we will call a "1-zero"
\eq\label{1zero}
\textrm{1-zero: }c_{1i}=0, \textrm{ for } i=\{3,4,\ldots,n-1\}\,.
\eqe
Other similar zeros are cyclic permutations of the above, $c_{2i}{=}0$, $c_{3i}{=}0$, etc. 
More generally, a "$k$-zero" has a form
\eq\label{2zero}
\textrm{$k$-zero: }c_{1i}=c_{2i}=\ldots c_{ki}=0, \textrm{ for } i\neq k{+}1,n\,,
\eqe
and similar for cyclic permutations. Example 1- and 2-zeros are highlighted in Figure \ref{fig:1}. A $k$-zero sets a number of $k(n{-}k{-}2)$ non-planar $c_{ij}$ to zero. All the various zeros are all independent, no zero implies any other zero.

For example, the ordered 4-point $\mathrm{Tr}(\phi^3)$ amplitude
\eq\label{4pt}
A_4(1234)=\frac{1}{s_{12}}+\frac{1}{s_{14}}=\frac{1}{s_{12}}-\frac{1}{s_{12}+s_{13}}\,, 
\eqe
clearly vanishes when imposing the 1-zero $c_{13}{=}{-}s_{13}{=}0$. This observation generalizes to arbitrary multiplicity, and identical facts hold for NLSM, YM-scalar, which unlike Tr$(\phi^3)$ also contains non-trivial numerators, as well as for YM after suitably including polarization vectors, as will be described later.

A related property is the splitting behavior. When setting the $c_{ij}$ to zero, except one invariant $c_{\star}$, the amplitude "splits" into three pieces: 
\eq
\mathcal{A}_n\left(c_{\star} \neq 0\right)=\left(\frac{c_{\star}}{X_{\mathrm{B}}X_{\mathrm{T}}}\right) \times \mathcal{A}^{\text {down }} \times \mathcal{A}^{\mathrm{up}}\,,
\eqe
tantalizingly similar to eq.(\ref{res}), but completely distinct physically, since $c_\star$ is not a pole of the amplitude, but quite the opposite, it leads to a zero. We will not use this further, and so leave the details to  \cite{Arkani-Hamed:2023swr}. 

In the same reference it was further conjectured that Tr$(\phi^3)$ amplitudes are in fact uniquely fixed by demanding a sufficient number of zeros. Consider again the simple 4-point example in eq.(\ref{4pt}), but now let the two terms have arbitrary coefficients. Imposing the zero condition $A|_{c_{13}=0}{=}0$ forces the two terms to be related, therefore fixing the amplitude up to some overall rescaling. Proving this conjecture for arbitrary multiplicity is one of the goals of this Letter. For simplicity we will mostly focus on the 1-zeros, but the general strategy should work for all zero types, as we discuss in Appendix C.

\section{Zeros vs BCFW shifts}
To demonstrate the proposed equivalence between zeros and BCFW scaling, we must first establish this connection at the level of kinematic data. First, each $k$-zero turns out to have a natural corresponding BCFW shift. For instance, for a 1-zero like $c_{1i}{=}0$, $i{\neq} 2,n$, the corresponding shift is $p_2{\rightarrow}p_2{+}z q$ and $p_n{\rightarrow}p_n{-}z q$, precisely the legs not involved in this zero. Similarly, for a 2-zero $c_{1i}{=}c_{2i}{=}0$, $i{\neq }3,n$, the corresponding shift is on $p_3$ and $p_n$, and so on. 

To see how the zero and shift may be related, let us write out the $X_{ij}$ in terms of $s_{ij}$. Consider the 1-zero type, $c_{1i}{=}0$. It is clear that the only $X_{ij}$ affected by either the related shift or the zero are of three possible types:
\begin{align}\label{zeroor}
\nonumber X_{1,j+1}&=s_{1,2,\ldots, j}= s_{12}+\sum_{i=3}^j s_{1i}+\sum_{i=3}^js_{2i}+s_{3\ldots j}\,,\\
\nonumber X_{2,j+1}&=s_{2,\ldots,j}= \sum_{i=3}^js_{2i}+s_{3\ldots j}\,, \\
X_{2,n}&=s_{2,\ldots,n-1}= -s_{12}-\sum_{i=3}^{n-1} s_{1i}\,,
\end{align}
for $2\le j\le n-2$. Under the zero condition, these become
\eqa\label{zeroeq}
\nonumber X_{1,j+1}^{(0)}&=& s_{12}+\sum_{i=3}^js_{2i}+s_{3\ldots j}\,,\\
\nonumber X_{2,j+1}^{(0)}&=& \sum_{i=3}^js_{2i}+s_{3\ldots j}\,,\\
X_{2,n}^{(0)}&=& -s_{12}\,,
\eqae
while the leading order in $z$ under a BCFW shift is given by
\eqa\label{zerobc}
\nonumber X_{1,j+1}^{(\infty)}&=& q\!\cdot\! p_1+\sum_{i=3}^jq\!\cdot\! p_i\,,\\
\nonumber X_{2,j+1}^{(\infty)}&=& \sum_{i=3}^jq\!\cdot \!p_i\,, \\
X_{2,n}^{(\infty)}&=& -q\!\cdot\! p_1\,.
\eqae
We now reach an important conclusion: as can be clearly seen from the above, the $X^{(0)}$ and the $X^{(\infty)}$ satisfy  the same $(n{-}3)$ linear relations (one for each $2{\le} j{\le} n{-}2$), corresponding to $c_{ij}{=}0$ in eq.(\ref{cij}). This connection between the $X^{(0)}$ and the $X^{(\infty)}$ remains true for all zero types, which we prove in Appendix A. In summary, we find a non-trivial connection between 
\begin{align}\label{cor}
&\nonumber \textrm{A general $k$-zero: } c_{ij}{=}0, \textrm{ for }\\
\nonumber &\quad i{=}\{1,2,\dots,k\}, j{=}\{k+2,k+3,\ldots,n-1\}\\
\nonumber &\textrm{and}\\
&\textrm{A BCFW shift: $p_{k+1}{\rightarrow} p_{k+1}{+}z q$, $p_n{\rightarrow} p_n{-}z q$}
\end{align}

The above can be cyclically permuted leading to similar statements for all zeros. In the next section we explore the connection more concretely.

\section{Subset zeros and UV scaling}
Here we will prove that any rational function built from planar invariants $X_{ij}$ satisfies a zero condition if and only if it also satisfies both a "subset enhanced UV scaling", and a "subset zero", as we explain below. 

Take $B$ any homogenous rational function of $X_{ij}$, which we established form a complete basis of invariants. Collect in subsets $B_i$ all individual terms of $B$ that under the shift $p_{k+1}{\rightarrow} p_{k+1}{+}z q$, $p_n{\rightarrow} p_n{-}z q$ scale as $z^i$ for $z{\rightarrow \infty}$. Then we can write
\eq\label{B}
B=B_m+B_{m-1}+B_{m-2}+\ldots\,,
\eqe
if we assume $z^m$ is the largest scaling for any term in $B$. Note the above is not a series expansion, but merely a partitioning of terms. We will denote the large $z$ behavior under BCFW shifts as $B|_{\textrm{UV}}$, so that originally $B|_{\textrm{UV}}{\sim} z^m$ and each $B_i|_{\textrm{UV}}{\sim} z^i$. For any zero and its corresponding BCFW shift, as described in eq.(\ref{cor}), we will show the following statements are equivalent
\eqa \label{c1}
\nonumber &&1.\ B \textrm{ satisfies the $k$-zero} \\
\nonumber &&2.\ \textrm{Each $B_i$ satisfies the $k$-zero}\\
 &&3.\ \textrm{Each $B_i$  has an enhanced UV scaling $z^{i-1}$}
\eqae
where by enhanced UV or BCFW scaling we mean that while individual terms in $B_i$ may scale  as $z^i$, they combine to produce an "enhanced" $z^{i-1}$ scaling. We call this novel type of UV scaling the "secret" or "subset" enhanced scaling. This set of equivalences is the first of our main claims.

Let us first prove the following fact: if a function $B$ satisfies a zero it must have enhanced BCFW scaling.
First, what does it mean for B to vanish under a zero? If we denote $X^{(0)}{\equiv} X|_{\textrm{zero}}$ as before, this simply means  $B(X_{ij}^{(0)}){=}0$. Since the only identities the $X_{ij}^{(0)}$ satisfy are given by $c_{ij}{=}0$, this means $B$ must be at least linear in the $c_{ij}$ that are being set to zero:
\eq\label{zer}
B=\sum_{c_{ij}\in
\mathrm{zero}} c_{ij}\{\ldots\}_{ij}\,.
\eqe
Next, what does it mean for  $B$ to have an enhanced scaling? Imposing the corresponding BCFW shift on eq.(\ref{B}) and taking the large $z$ limit we find
\eq
B\rightarrow z^mB_m(X_{ij}^{(\infty)})+\mathcal{O}(z^{m-1})\,,
\eqe
since we assumed all terms in $B_m$ scale as $z^m$. Enhanced scaling simply means $B_m(X_{ij}^{(\infty)}){=}0$. But this can only occur if
\eq
B_m=\sum_{c_{ij}\in
\mathrm{zero}} c_{ij}\{\ldots\}_{ij}\,,
\eqe
with the same $c_{ij}$ that appear in eq.(\ref{zer}), since we established these are the only identities that the  $X_{ij}^{(\infty)}$ satisfy. The two conditions are identical, so we conclude that if $B$ satisfies a zero it must also  have enhanced scaling. But we can go further. If $B_m$ has the above form, this directly implies $B_m$ also satisfies the zero on its own, $B_m(X_{ij}^{(0)}){=}0$. But this in turn implies that
\eq
B'=B_{m-1}+B_{m-2}+\ldots\,,
\eqe
also satisfies the zero. We can now apply the same reasoning starting from $B'$, proving that $B_{m-1}$ also satisfies the zero and has enhanced scaling, and so on, proving the equivalences in (\ref{c1}). 

Therefore, we have shown that for general functions of $X_{ij}$, the zero condition is equivalent to both each subset (defined by the individual scaling of terms) satisfying the zero independently, and also  each subset having  enhanced BCFW scaling. This observation is curiously similar to the fact that permutation invariance also always implies an enhancement of the BCFW scaling from any $z^{\textrm{odd}}$ to $z^{\textrm{even}}$ \cite{McGady:2014lqa}.

In particular, these results hold for all ordered amplitudes, such as Tr($\phi^3$), NLSM, or YM-scalar. In the next section we exploit this type of reasoning to the concrete case of Tr($\phi^3$), in order to prove it is uniquely fixed by a sufficient number of distinct zero conditions.

\section{Tr($\phi^3$) subset zeros and uniqueness}
We now specialize to the case where our function $B$ is a general local, ordered, ansatz for Tr($\phi^3$) amplitudes. In this case we assume all ansatz terms have numerators which do not carry any momentum dependence, and denominators which can be associated to cubic tree graphs. Or put more simply, the ansatz is given by the sum of all Feynman diagrams of Tr($\phi^3$) theory, but with arbitrary coefficients.

In the previous section we saw that the zero for any function can be decomposed into zeros of individual subsets, organized by their UV scaling. In our present case, these subsets can be further decomposed into what we call $D$-subsets, which have a simple meaning in terms of graph topologies. Let us focus on the simplest case of a 1-zero, $c_{1i}{=}0$.

\begin{figure}[h] 
   \centering
   \includegraphics[width=2.3in]{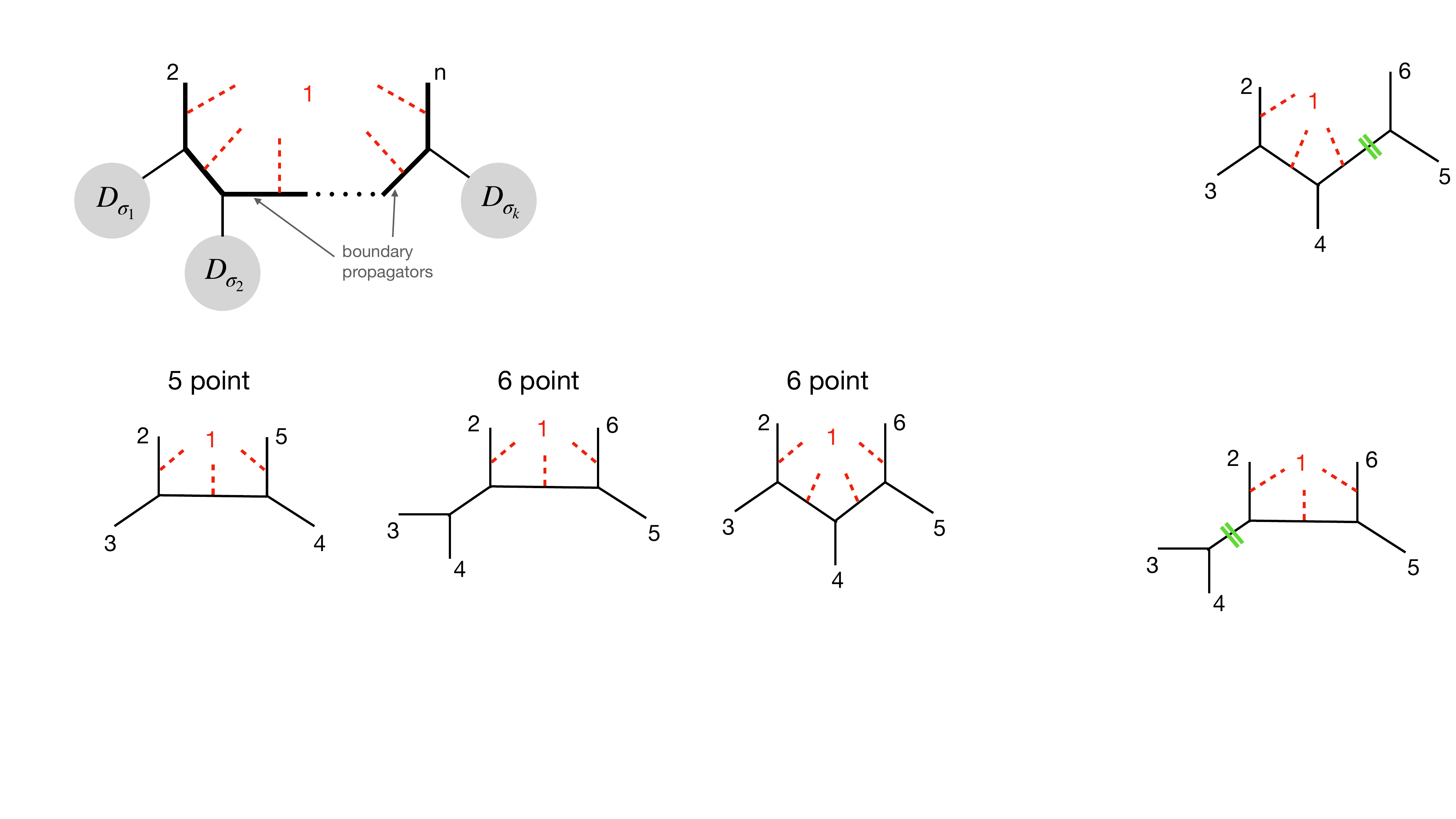} 
   \caption{Obtaining higher point diagrams from lower point. Leg 1 can be attached in any position marked by the red lines.}  \label{1}
\end{figure}

To generate the  $D$-subsets at $n$-point for this zero, start with an $(n{-}1)$-point diagram, with leg 1 missing. Then the $D$-subset corresponding to this $(n{-}1)$-point diagram is formed by adding leg 1 in all possible ways (that respect ordering and trivalent interactions). This is shown in Figure \ref{1}. Here note that what we call  "boundary propagators" are precisely the $X_{ij}$ in eq.(\ref{zeroor}), that we found participate in either the zero conditions or attain $z$-dependence under shifts. Meanwhile, the $D_{\sigma_i}$ contain the remaining subgraphs. All $n$-point diagrams can be uniquely obtained via this procedure by starting from all $(n{-}1)$-point diagrams, for all possible choices of arranging the propagators and external legs in the $D_{\sigma_i}$. Taking a general linear combination of all such terms gives us the local ansatz for Tr($\phi^3$) amplitudes.

Let us start with an example at 6-point. There are a total of 14 diagrams, and 5 different subsets having either 1, 2, or 3 boundary propagators. We will illustrate the subset with two boundary propagators, which split into two $D$-subsets, as shown below
   \includegraphics[width=3.3in]{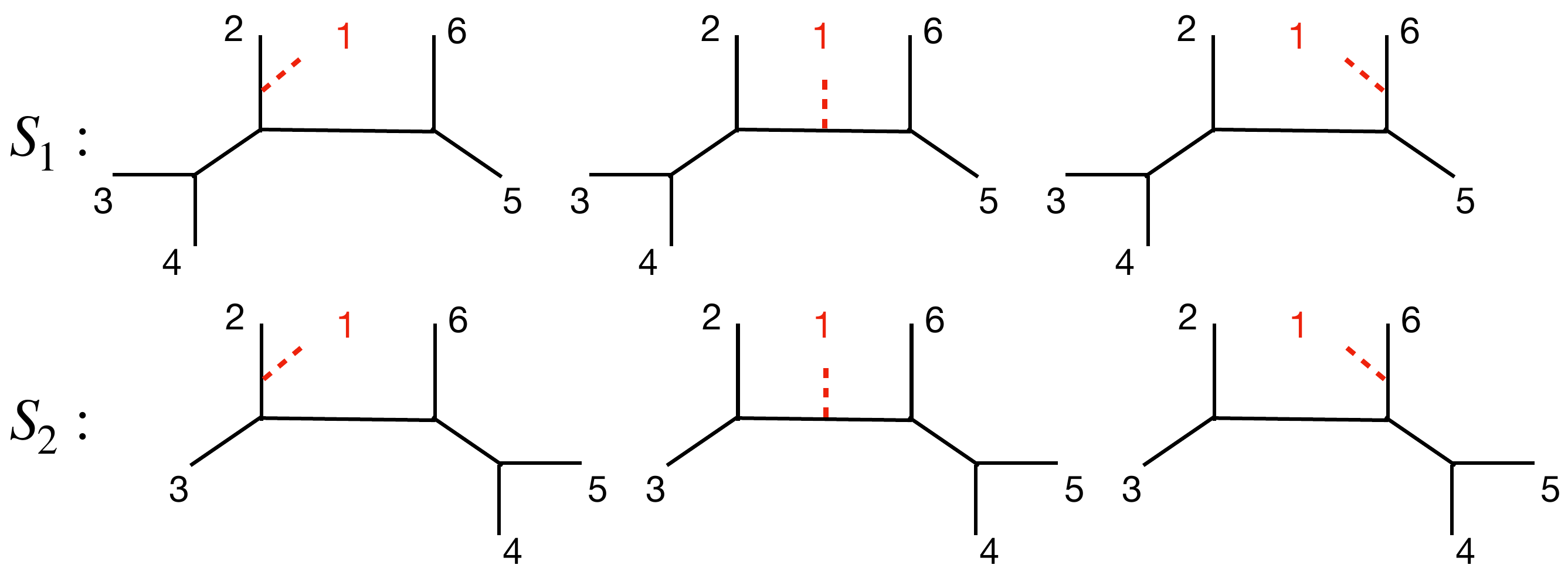}  
Explicitly the $D$-subsets  are
\eqa\label{dsub}
\nonumber S_1&=&\frac{x_1}{s_{12}s_{1234}s_{34}}+\frac{x_2}{s_{234}s_{1234}s_{34}}+\frac{x_3}{s_{234}s_{2345}s_{34}}\,,\\
S_2&=&\frac{x_4}{s_{12}s_{123}s_{45}}+\frac{x_5}{s_{23}s_{123}s_{45}}+\frac{x_6}{s_{23}s_{2345}s_{45}}\,.
\eqae
First, each of the 6 terms above individually scales as $z^{-2}$, since they all contain two boundary propagators. Using the results of the previous section, we directly know that if the 6-point amplitude satisfies a zero $c_{1i}{=}0$, then we also must have $\left(S_1{+}S_2\right)|_{\textrm{zero}}{=}0$, and $\left(S_1{+}S_2\right)|_{\textrm{UV}}{\sim} z^{-3} $.

However, we can prove three stronger statements:
\begin{enumerate}
\item All $D$-subsets independently satisfy the 1-zero (and have enhanced UV scaling under the ($p_2$, $p_n$) shift)
\item All $D$-subsets are uniquely fixed up to an overall coefficient
\item Imposing $(n{-}3)$ distinct 1-zeros  uniquely fixes the ansatz as the $n$-point Tr($\phi^3$) amplitude, up to an overall coefficient
\end{enumerate}

{\bf 1. All $D$-subsets independently satisfy the zero}
Consider the $D$-subsets in eq.(\ref{dsub}), which satisfy $(S_1{+}S_2)|_\textrm{zero}{=}0$. Next, cut the non-boundary propagator $s_{34}$, which uniquely picks the diagrams in $S_1$:
\eq
(S_1{+}S_2)|_\textrm{zero}|_{s_{34}-\textrm{cut}}{=}S_1|_\textrm{zero}|_{s_{34}-\textrm{cut}}=0\,,
\eqe
so $S_1$ must have a form
\eq
S_1=\frac{c_{13}\{\ldots\}+c_{14}\{\ldots\}+ c_{15}\{\ldots\}+s_{34}Q}{s_{34}(s_{12}s_{234}s_{1234}s_{2345})}\,,
\eqe
it is clear from eqs.(\ref{zeroor}-\ref{zeroeq}) that no linear combination of the boundary propagators $s_{12},s_{234},s_{1234}, s_{2345}$ can create the non-boundary propagator $s_{34}$. Therefore we must have $Q{=}0$, and so $S_1|_\textrm{zero}{=}0$ even away from the cut. The argument can be repeated for the $D$-subset $S_2$, by cutting $s_{45}$, and similarly for all other $D$-subsets. This implies
\eq
\nonumber \textrm{\emph{Each $D$-subset vanishes under the 1-zero. }}
\eqe
and so by the previous section, 
\eq
\nonumber \textrm{\emph{Each $D$-subset also has enhanced UV scaling}}
\eqe

{\bf 2. $D$-subsets are uniquely fixed}
We will show the zero condition implies all relative coefficients of any $D$-subset are uniquely fixed. Consider the last two diagrams in the set $S_1$ in eq.(\ref{dsub}), and cut their common propagator $s_{234}$, implying
\eq\label{subsetx}
\left.\left.S_1\right|_{\textrm{zero}}\right|_{s_{234}-\textrm{cut}}=\left.\left.\left(\frac{x_2}{s_{1234}}+\frac{x_3}{s_{2345}}\right)\right|_{\textrm{zero}}\right|_{s_{234}-\textrm{cut}}=0\,.
\eqe
We can write the two propagators as
\eqa
\nonumber s_{1234}&=&s_{12}+s_{13}+s_{14}+s_{234}\,,\\
s_{2345}&=&-(s_{12}+s_{13}+s_{14}+s_{15})\,,
\eqae
so, under the zero condition $c_{1i}{=}0$ and the additional cut $s_{234}{=}0$ we find $s_{1234}{=}{-}s_{2345}$, and eq.(\ref{subsetx}) can only hold if $x_2{=}x_3$. Repeating this for the first pair of diagrams, by cutting the common propagator $s_{1234}$, we obtain $x_1{=}x_2$, proving that all terms in this $D$-subset have equal coefficients. The whole process can be carried out for each $D$-subset, so we conclude 
\eq
\nonumber \textrm{\emph{$D$-subsets are fixed uniquely by the 1-zero condition}}
\eqe

{\bf3. Amplitude uniqueness}
Finally we can prove that imposing $(n{-}3)$ distinct 1-zero conditions is sufficient to fully determine the Tr$(\phi^3)$ amplitude. We can illustrate uniqueness by considering a graph whose vertices are represented by the diagrams. Let two diagrams be connected in the graph if there exists a zero condition which places them in the same subset, implying the two "adjacent" diagrams have equal coefficient. If we can show the  graph of all diagrams is connected, this implies any two diagrams have equal coefficient, so the amplitude is fixed up to an overall number. 
This can  be proven by induction: consider two distinct $n$-point diagrams. If they are not already part of the same subset for some zero, remove a leg. Repeat this until they become diagrams  in the same subset. In the worst-case scenario this can go on at most until reaching the 3-point diagram, so no two diagrams may be more than $n{-}3$ steps away from each other. This proves one of our main claims, that\\

\emph{Starting from a local ansatz with trivial numerators, $(n{-}3)$ 1-zeros uniquely fix the} Tr\emph{$(\phi^3)$ amplitude, up to overall normalization.}\\

This result holds for all zero types, but we do not pursue a general proof. We present how the argument generalizes for the 2-zero case in Appendix C. 

\section{Yang-Mills polarization structures}

It is natural to ask how constraining the amplitude zero is for YM amplitudes. In this case we can write cubic, ordered, local ansatze with numerators of mass dimension $[n{-}2]$, multilinear in polarization vectors $e_i$ which satisfy $e_i{\cdot p_i}{=}0$. At 4-point such an ansatz looks like
\eq\label{YM}
B_4=x_1\frac{e_1\!\cdot\! e_2 e_3\!\cdot\! e_4 p_1\!\cdot\! p_3}{p_1\!\cdot\! p_2}+x_2\frac{e_1\!\cdot\!  e_2 e_3\!\cdot\!  p_1 e_4\!\cdot\! p_3}{p_1\!\cdot\!  p_2}+\ldots\,.
\eqe
Note the two terms above have different polarization structure: the first term has 0 factors of $e{\cdot} p$, while the second has two such terms. These are what we call the distinct polarization structures, and an $n$-point amplitude has $\lfloor n/2\rfloor$ such structures.

For YM, the zero condition requires only a trivial extension to include polarization vectors:
\eq\label{ym}
e_i\!\cdot\! e_j= e_i\!\cdot\! p_j= p_i\!\cdot\! p_j=0, \textrm{ for {$j$} not adjacent to {$i$}}\,,
\eqe
and where $i$ as above can be any set $\{1\},\{1,2\},\ldots$, plus cyclic permutations, similarly to the usual scalar zeros. For the 4-point example above, the zeros we use are $e_1{\cdot}e_3{=}e_1{\cdot}p_3{=}p_1{\cdot}e_3{=}p_1{\cdot}p_3{=}0$, plus cyclic. From the nature of the zero constraint, it is immediately clear zeros cannot fix the full amplitude: no relation  can be obtained between the two terms in eq.(\ref{YM}). It is however interesting to combine with constraints due to requiring the BCJ relations to hold, which require $\sum_{i=2}^{n-1} k_{1i}A(2,\ldots,i,1,i+1,\ldots,n){=}0$, where $k_{1i}{=}\sum_{j=2}^i s_{1j}$.  

Remarkably, we find that up to 6-point, combining the two conditions allows precisely $[n/2]$ solutions, corresponding to the different tensor structures! So while the YM amplitudes are not fully fixed, they are highly constrained, and it would be interesting to combine this observation with general BCJ bootstrap procedures \cite{Carrasco:2019yyn,Carrasco:2021ptp,Pavao:2022kog}. We give the explicit formulas for these polarization structures at 4-point in Appendix D.
 
\section{Outlook}
In this work we showed hidden amplitude zeros are in fact closely related to UV scaling, a crucial ingredient of the BCFW recursion. Beyond this connection, the similarities between zeros and the splitting behavior on the one hand, and BCFW scaling, recursion, and unitarity on the other, suggest some deeper structure exists, that may lead to a novel understanding (and new computational methods) for scattering amplitudes. An important related caveat of our claim that unitarity emerges from zeros is that the uniqueness cannot relate the normalization of amplitudes at different multiplicities. Some extra input is therefore required, with the related splitting behavior near zeros being a natural candidate. 

The approach we developed can be extended to prove uniqueness in much more general cases (including beyond assuming locality, for other theories like NLSM, YM and YM-scalar, as well as to loop level for the first time), directions currently under investigation \cite{prog}. This is motivated by the new "surface kinematics" \cite{Arkani-Hamed:2024tzl} that for the first time provides a canonical definition for loop integrands, which are rational functions and in some cases may also be uniquely determined by zeros. \\

\noindent {\it Acknowledgments:} The author would like to thank Nima Arkani-Hamed, Jeffrey Backus, Carolina Figueiredo and Song He for valuable discussions. This Letter is supported by the Beijing Natural Science Foundation International Scientist Project No. IS24014, and the National Natural Science Foundation of China General Program No. 12475070.


\appendix
\section{Appendix A: UV scaling vs general zeros}
In this appendix we prove that all zeros are equivalent to enhanced scaling under some corresponding BCFW shift. We accomplish this by showing that the $X_{ij}^{(0)}$ and $X_{ij}^{(\infty)}$ satisfy the same sets of identities, defined by $c_{ij}{=}0$.
First, we already established that the zeros $c_{ij}{=}0$ induce linear relations between the $X_{ij}$. The set of all $X_{ij}$ that attain such dependencies can be easily obtained from the definition eq.(\ref{cij}). For zero a $c_{ij}{=}0$, for $i{=}{1,2,\ldots,a{-}1}$, this set is given by all $X_{ij}$ with $i{\le}  a$ and $ a{+}1{\le} j$. This is a total of $(k{+}1)(n{-}k{-}1){-}2$ elements. The remaining $X_{ij}$ are independent under the zero condition. These sets can be visualized in the kinematic mesh, as the vertices of squares participating in the zero.

Next, we are interested in which $X_{ij}$ attain $z$ dependence under a BCFW shift acting on particles $a$ and $n$. Since $X_{ij}{=}(p_i{+}p_{i+1}{+}\ldots{+}p_{j-1})^2$, for $X_{ij}$ to have dependence on $z$, we must have $i{\le} a{\le} j{-}1$, otherwise the $z$ shift cancels out. This is precisely the same set of $X_{ij}$ we found above for the zero condition! 

\begin{figure}[htbp] 
   \centering
   \includegraphics[width=2.8in]{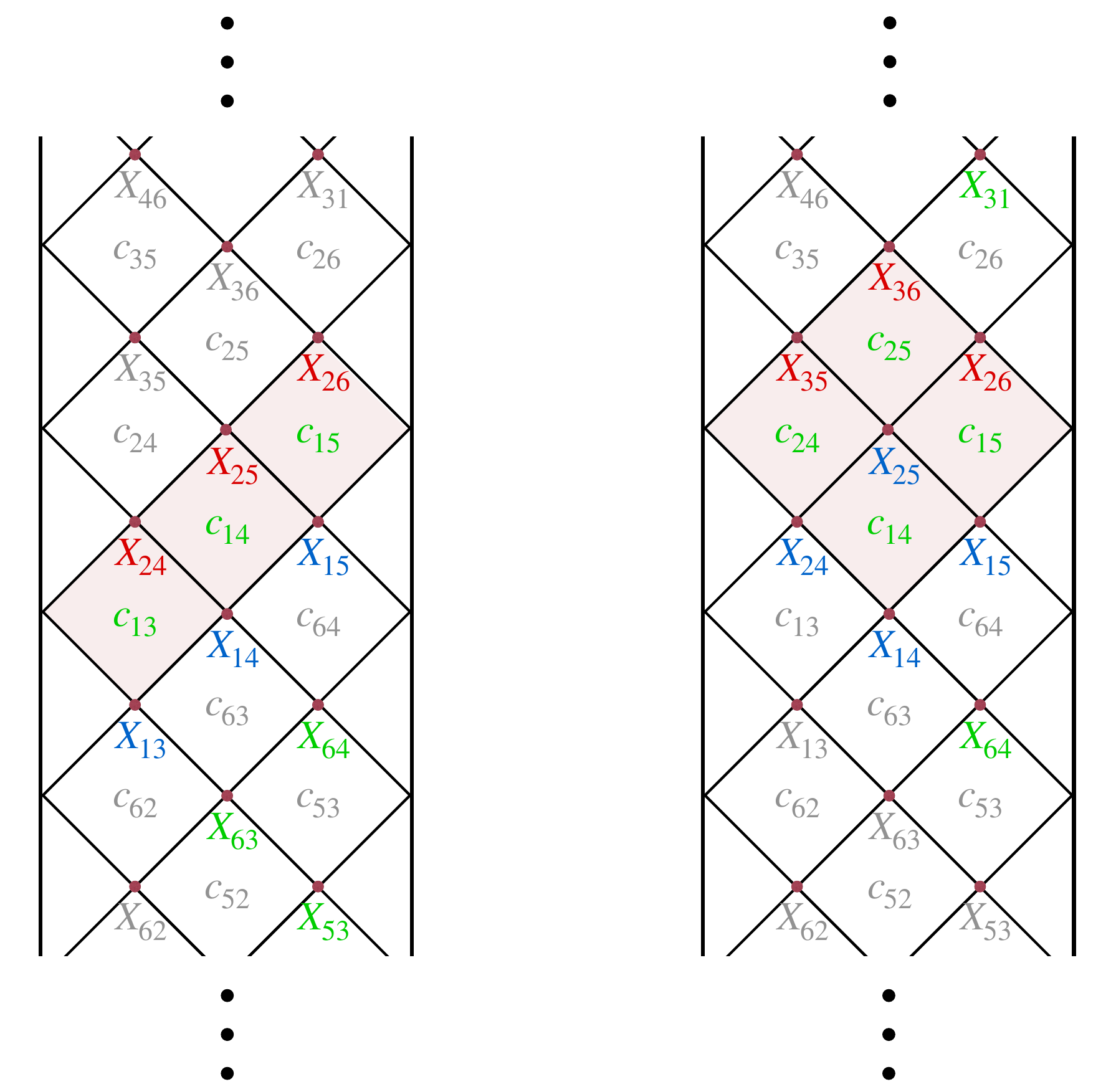} 
   \caption{A kinematic 6-point basis can be obtained by replacing some of the $X_{ij}$ (shown in blue) in the original basis with the corresponding $c_{ij}$ that are part of the zero (shown in green). The only $z$-dependent basis elements are those shown in red. The basis elements in green are all $z$-independent.}
   \label{fig:3}
\end{figure}

Finally we need to find all linear combinations of $z$-dependent $X_{ij}$, in which the $z$-dependence cancels out. For this we go to a mixed basis of $c_{ij}$ and $X_{ij}$. For a given $k$-zero with a set of $c_{ij}{=}0$, we can form a basis by replacing $k(n{-}k{-}2)$ planar invariants $X_{i_1,i_2}$ by the $k(n{-}k{-}2)$ corresponding non-planar invariants $c_{i_1,i_2}$ that appear in the zero set. In such a basis, the only $z$-dependence can come from the $(n{-}3)$  $X_{ij}$ terms on the top edges of the $k$-zero rectangle, shown in red in Figure \ref{fig:3}. We need to show there are no linear combinations between these $X_{ij}$ where the $z$-dependence drops out.
For a $k$-zero, the three types of such $X_{ij}$ are given by
\eqa
\nonumber X_{k+1-i,n}^{(\infty)}&=&\sum_{l=k+1-i}^{n-1} q\cdot p_l, \quad i\ge 1\,, \\
\nonumber X_{k+1,n-j}^{(\infty)}&=&\sum_{l=k+2}^{n-1-j} q\cdot p_l, \quad j\ge 1\,, \\
X_{k+1,n}^{(\infty)}&=&\sum_{l=k+2}^{n-1} q\cdot p_l \,,
\eqae
and it is clear that all terms are linearly independent. This shows that the only relations satisfied by the $X_{ij}^{(\infty)}$ are those corresponding to $c_{ij}{=}0$, and so  $X_{ij}^{(\infty)}$ and  $X_{ij}^{(0)}$ satisfy identical relations.

\section{Appendix B: $N$-point proof for subset uniqueness in Tr$(\phi^3)$ }
\begin{figure}[h] 
   \centering
   \includegraphics[width=3.4in]{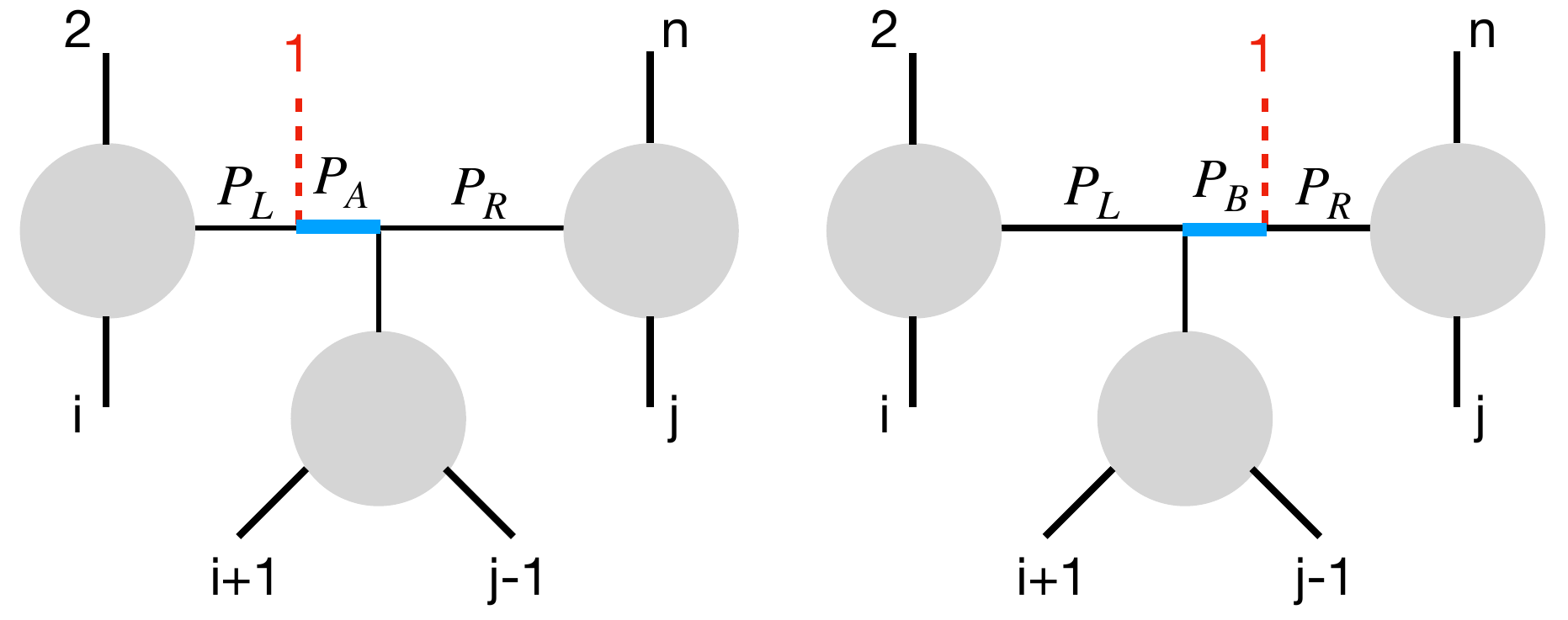} 
   \caption{Two neighbor diagrams differ only by the blue propagator.}
   \label{fig3}
\end{figure}

In this appendix we prove that all $D$-subsets defined for Tr$(\phi^3)$ are uniquely fixed by requiring the amplitude satisfy a zero, ie. that any two neighbor diagrams in a given $D$-subset have equal coefficient. As before, we consider two diagrams in this subset which differ by exactly one boundary propagator, as shown in Figure \ref{fig3}.
These have a form
\eq
\frac{1}{P_L P_R}\left(\frac{x_i}{ P_A}+\frac{x_{i+1}}{ P_B}\right)\,.
\eqe
We need to prove $P_A{=}{-}P_B$ on the amplitude zero plus ($P_L, P_R$)-cut locus. We can write
\eqa
P_A&=&s_{12}+\sum_{k=3}^i s_{1k}+P_L\,,\\
P_R&=&s_{12}+\sum_{k=3}^{j-1}s_{1k}+P_B\,,
\eqae 
so, under the zero condition, and on the cut $P_L{=}P_R{=}0$, we find $P_A{=}-P_B$ and conclude the two diagrams  cancel if they have equal coefficients, $x_i{=}x_{i+1}$. This can be repeated for all pairs in the subset, and then for all subsets, proving that one amplitude zero uniquely fixes all subsets. The final step is to combine this result with constraints from distinct zero conditions on other legs, which as discussed in the main text implies the full amplitude is indeed uniquely fixed after imposing some sufficient number of zeros.

\section{Appendix C: Tr($\phi^3$) uniqueness from general zeros}
In this Appendix we describe in  more detail the case for a 2-zero $c_{1i}{=}c_{2i}{=}0$, $i{\neq} 3,n$, to demonstrate how the uniqueness proof generalizes. For this 2-zero, the relevant $D$-subsets are obtained in a similar way, as all possible ways to add both legs 1 and 2 to an $n{-}2$ point diagram. Note that adding the two legs can be done in two different ways - either as separate legs, or as a two-particle pole $1/s_{12}$. So, for instance, by starting with one of the 4-point diagrams, we obtain two $D$-subsets as shown below

   \includegraphics[width=3.2in]{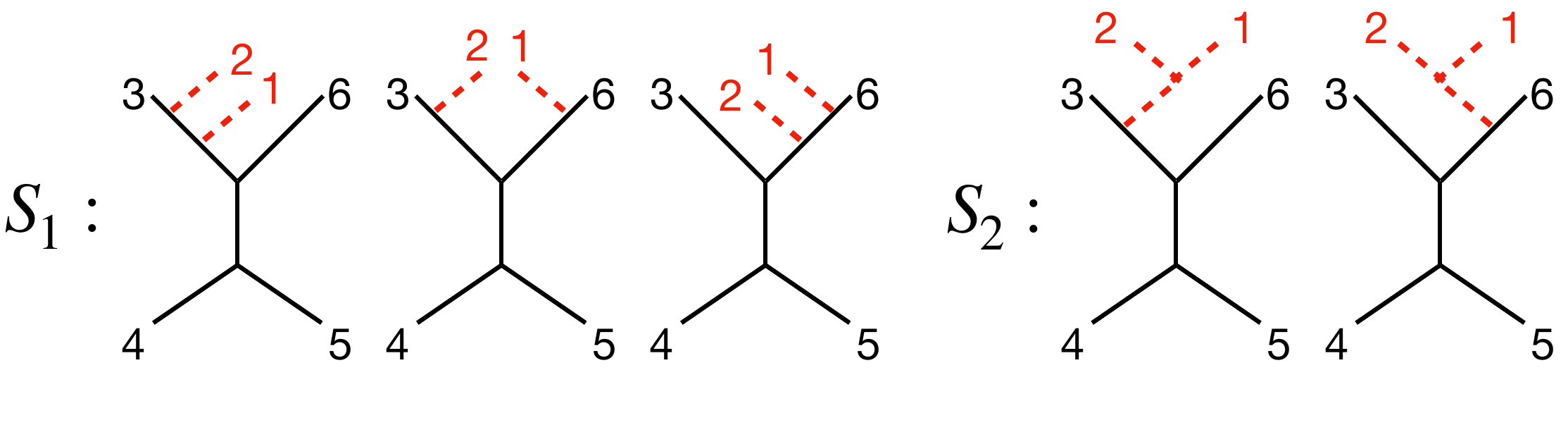} 
Given by
\eqa
\nonumber S_1&=&\frac{1}{s_{45}}\left(\frac{x_1}{s_{23}s_{123}}+\frac{x_2}{s_{23}s_{2345}}+\frac{x_3}{s_{345}s_{2345}}\right)\\
S_2&=&\frac{1}{s_{45}s_{12}}\left(\frac{x_4}{s_{123}}+\frac{x_5}{s_{345}}\right)\,.
\eqae

 Note that in this case, the propagator $s_{12}$ is not a boundary propagator, as it is not affected by the 2-zero via eq.(\ref{cij}) (or more easily, by our previous results this also follows because $s_{12}$ is invariant under the corresponding shift, which is on $p_3,p_n$). We do not pursue a general proof in more detail, but it should be clear the strategy in the main text should work directly: isolate subsets via cuts, show the cuts can always be chosen to select two neighbor diagrams, fixing their relative coefficient, and so on. Conceptually, this should work for all types of $k$-zeros.

\section{Appendix D}
We list below the two different YM polarization structures at 4-point, fixed by combining the zero and BCJ relations 
\begin{widetext}
\eqa
\nonumber A_4^{(0)}&=&p_1\!\cdot\! p_3 \left(\frac{e_1\!\cdot\! e_2 e_3\!\cdot\! e_4}{p_1\!\cdot\! p_2}-\frac{e_1\!\cdot\! e_4 e_2\!\cdot\! e_3}{p_1\!\cdot\! p_4\! }\right)+e_1\!\cdot\! e_3 e_2\!\cdot\! e_4\,\\
\nonumber A_4^{(2)}&=&\frac{1}{p_1.p_2}\biggl( e_3.e_4 e_1.p_3 e_2.p_1-e_1.e_4 e_2.p_1 e_3.p_1-e_1.e_4 e_2.p_1 e_3.p_2+e_1.e_3 e_2.p_1 e_4.p_1+e_1.e_2 e_3.p_2 e_4.p_1+e_1.e_3 e_2.p_1 e_4.p_2\\
&&-e_1.e_2 e_3.p_1 e_4.p_2-e_1.p_2 \left(e_3.e_4 e_2.p_3-e_2.e_4 \left(e_3.p_1+e_3.p_2\right)+e_2.e_3 \left(e_4.p_1+e_4.p_2\right)\right)\biggl)+\textrm{cyclic}\,.
\eqae
\end{widetext}

\bibliographystyle{apsrev4-1}

\end{document}